\documentclass[aps,prb,fleqn,twocolumn,showpacs,floatfix]{revtex4}   %Final preview
\usepackage{graphicx}   % Include figure files
\usepackage{dcolumn}    % Align table columns on decimal
\usepackage{bm}     % bold math
\usepackage{color}  %Just for correction purpose

\begin{document}

\title{Semiconducting-enriched single wall carbon nanotube
networks \\ applied to field effect transistors}

\author{N. Izard, S. Kazaoui$^{*}$, K. Hata, T. Okazaki,
 T. Saito, S. Iijima, and N. Minami}
\affiliation{National Institute of Advanced Industrial Science and
Technology (AIST)\\1-1-1 Higashi, Tsukuba, Ibaraki, 305-8565
Japan}

%Abstract
\begin{abstract}
Substantial progress are reported on field effect transistors
(FET) consisting of semiconducting single wall carbon nanotubes
(s-SWNTs) without detectable traces of metallic nanotubes and
impurities. This outstanding result was made possible in
particular by ultracentrifugation (250,000g) of solutions composed
of SWNT powders with polyfluorene as extracting agent in toluene.
Such s-SWNTs processable solutions were applied to realize FET,
embodying randomly or preferentially oriented nanotube networks
prepared by spin coating or dielectrophoresis. Devices exhibit a
stable p-type semiconductor behavior in air with very promising
characteristics: the on-off current ratio is $10^5$, the
on-current level is around 10~$\mu$A and the estimated hole
mobility is larger than 2~cm$^{2}$/vs. The present results are
demonstrated by optical absorption, Raman and electrical
measurements.

\end{abstract}
\maketitle

Novel semiconductors materials for field effect transistor FET and
thin-film transistors TFT are highly demanded. \cite{Review-FET}
In particular, semiconducting single wall carbon nanotubes
(s-SWNTs) are very promising, because individual s-SWNTs are known
to exhibit high on-off current ratio, high electron/hole mobility,
to carry high current density and to operate at high frequencies.
\cite{Biercuk, Avouris-FET, mobility, Nico-APL} Several groups
have demonstrated excellent transfer characteristics using
individual s-SWNTs, \cite{Biercuk, Avouris-FET, mobility,
Nico-APL} but generally poor ones using ensemble (network and thin
film) of SWNTs due to traces of metallic nanotubes (m-SWNTs) and
impurities (catalytic and amorphous particles). \cite{Snow-APL,
Arnold-NN, Fukao} Therefore, very efficient methods to selectively
synthesis s-SWNTs or to selectively extract s-SWNTs from as-grown
nanotubes are still required.

In recent years, several approaches to extract s-SWNTs from
nanotube powders were explored using for instance chemical
functionalization, \cite{Strano, Cecilia-JACS, Maeda-JACS} DNA and
polymers wrapping, \cite{DNA, kaza} and density gradient
ultra-centrifugation techniques.\cite{Arnold-NN} The latter
efficiently separates s-SWNTs and m-SWNTs, but traces of
surfactant and density gradient materials limit the performances
of the FET. Very recently, two groups have reported on the
selective extraction of near-armchair s-SWNTs from nanotube
powders using polyfluorene as extracting agent. \cite{Nish-NN,
Chen-NL} According to Nish \textit{et al.}, \cite{Nish-NN} the
sample shows no detectable traces of m-SWNTs based only on optical
spectroscopy. However, neither the electrical properties nor the
fabrication of FET devices were addressed.

\begin{figure}
\includegraphics[width=7.5cm, clip]{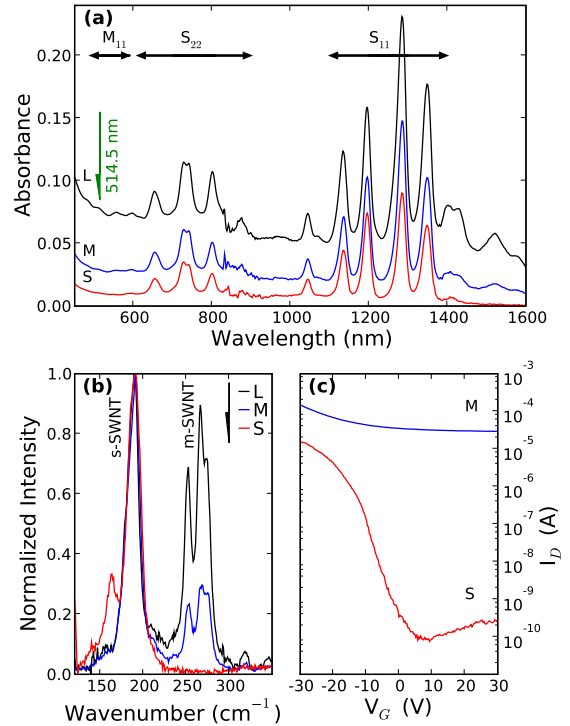}
\caption{(Color online): (a) shows the optical absorption spectra
and (b) the Raman spectra at 514.5 nm of L, M and S samples . (c)
displays the transfer characteristic, I$_D$ vs V$_G$ for
V$_{DS}$=-14V of FET devices made of sample M and S.} \label{fig1}
\end{figure}

In this letter, we report on the electronic properties of FET
consisting of \textit{semiconducting-enriched single wall carbon
nanotubes, without detectable traces of metallic nanotubes and
impurities}, with in our detection limits. This unprecedented
achievement is made possible by ultracentrifugation (250,000g),
sonication and filtration of solutions composed of SWNT powders
with polyfluorene as extracting agent in toluene. Evidences are
gathered by optical absorption, Raman and electrical measurements
(see Fig.\ref{fig1}). We shall demonstrate that such s-SWNTs
realize high-performances FET devices compare to networks/thin
films of SWNTs and solution processable polymers/organic
materials. \cite{Review-FET, Arnold-NN, brutting}

S-SWNTs solutions were prepared as follow. First, SWNTs powders
(as-prepared HiPco, Carbon Nanotechnologies Inc.), PFO
(Poly-9,9-di-n-octyl-fluorenyl-2,7-diyl, Sigma-Aldrich) and
toluene were mixed in the following ratio SWNT~(5~mg):~PFO
(5~mg):~toluene~(30~ml) and homogenized by sonication (1 hour
using a water-bath and 5 minutes using a tip sonicators). Then,
this mixture was centrifugated for 5$\sim$120 minutes using either
a desktop centrifuge (angle rotor type) or an ultra-centrifuge
(swing rotor type). Next, the upper 80~\% of the supernatant
solution was collected. To recover only the SWNTs while washing
out the PFO, the supernatant solution was filtered through
0.1~$\mu$m Teflon filter and rinsed with toluene several time
(until the characteristic optical absorption band of PFO at 385nm
completely disappears from the filtrate). Finally, this filter was
soaked in organic solvent (such as Toluene or
N-Methyl-2-pyrrolidone (NMP)) and subjected to mild sonication. We
shall focus on three types of SWNTs processable solutions
centrifugated at 10000~g for 15~min (labeled L), 250000~g for
30~min (labeled M) and 250000~g for 60~min (labeled S), with all
the other processing parameters identical. These samples were
characterized by optical absorption, Raman and electrical
measurements. To exclude residues of polymers and solvents, FET
devices were annealed at 300-400$^\circ$C for 1 hour in vacuum or
nitrogen. Note that the transmission electron microscopy (TEM)
reveals significant amount of catalytic particles in sample L but
no detectable traces in sample M and S.

\begin{figure}
\includegraphics[width=7.5cm, clip]{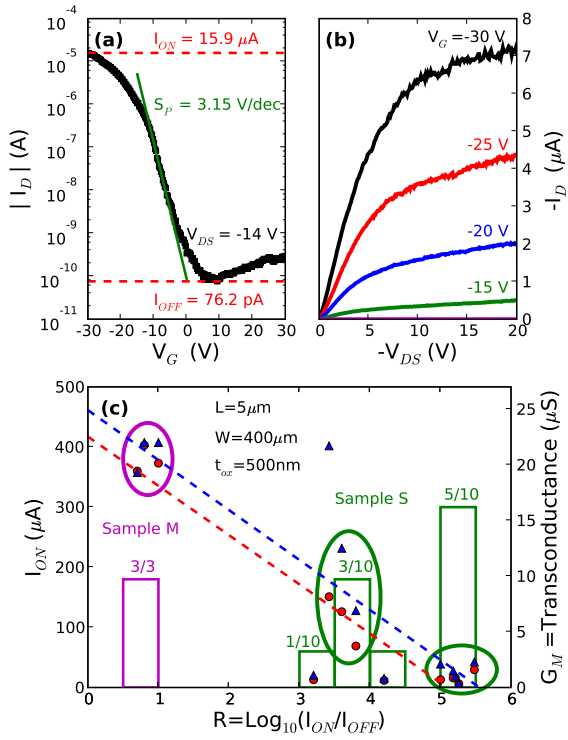}
\caption{(Color online): (a) and (b) display the transfer
characteristic of an FET device made by spin-coating. (c) presents
the histogram, $I_{ON}$  ($- \bigcirc -$) and Transconductance ($-
\bigtriangleup -$) versus $Log_{10}(I_{ON}/I_{OFF})$ for sample S
and M. Lines represent the linear fit} \label{fig2}
\end{figure}

\begin{figure}
\includegraphics[width=7.3cm, clip]{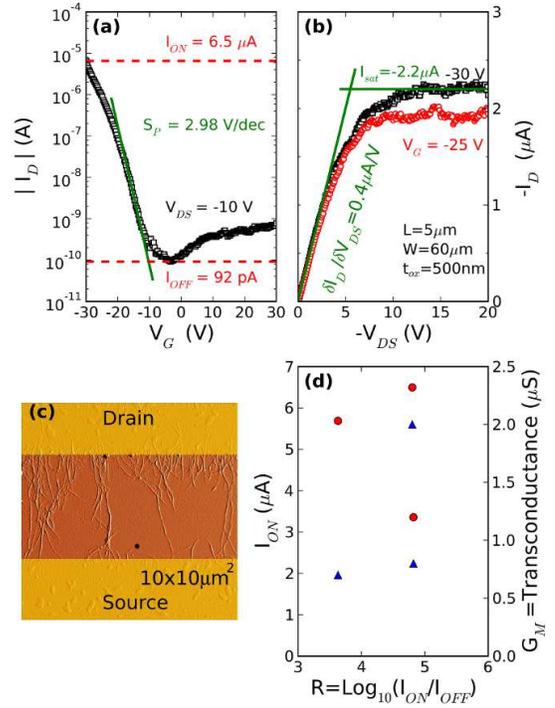}
\caption{(Color online): (a) and (b) show the transfer
characteristic of an FET device made by electrophoresis. (c)
displays the AFM image of a s-SWNTs network. (d) shows $I_{ON}$
($- \bigcirc -$) and Transconductance ($- \bigtriangleup -$)
versus $Log_{10}(I_{ON}/I_{OFF})$.} \label{fig3}
\end{figure}

Figure \ref{fig1}a displays the optical absorption spectra of L, M
and S samples in toluene. Sharp peaks in the range 1100-1400 and
600-900~nm, labeled S$_{11}$ and S$_{22}$, corresponds to the
first and to the second optical transitions in s-SWNTs,
respectively. Whereas, less resolved peaks in the range
500-600~nm, labeled M$_{11}$, are assigned to optical transitions
from m-SWNTs. The intensity of these features as well as the
absorption background decrease in the following sequence L, M and
S. It is worth noticing that the optical absorption spectrum of
sample S is dominated by S$_{11}$ and S$_{22}$ features (s-SWNTs)
without detectable traces of M$_{11}$ peaks (m-SWNTs). To
ascertain this result, the Raman spectra were recorded at
514.5~nm, which resonantly probes m-SWNTs. We observed that the
Radial Breathing Mode (RBM) around 270~cm$^{-1}$ (m-SWNTs)
normalized to the RBM mode around 190~cm$^{-1}$ (s-SWNTs)
decreases from L to M, and eventually vanishes in sample S (Fig.
\ref{fig1}b). It is essential to corroborated those results by
electrical measurements.

The electrical properties were investigated in field effect
configuration. A few droplets of SWNTs solutions were spin-coated
on Si/SiO$_2$ wafer, with pre-patterned drain and source
electrodes (Cr/Au) and Si as a back gate electrode. Figure
\ref{fig1}c present the drain current $I_{D}$ versus gate bias
$V_{G}$ with drain-source bias $V_{DS}$=-14~V. The sample S
presents a sizable gate bias dependence characteristic of a p-type
semiconductor behavior, whereas sample M exhibits a weaker gate
effect and higher conductivity typical of metallic behavior.
Similar results were observed at least on 10 devices made from
sample S and 3 devices from sample M (see the histogram in Fig.
\ref{fig2}c). The electrical measurements unambiguously prove that
sample S essentially consists of s-SWNTs, whereas sample M (and to
greater extent L, not shown) contain s-SWNTs, m-SWNTs and probably
impurities.

At this point it is important to stress that the electrical,
optical absorption and Raman measurements provide strong evidence
that sample S consist of s-SWNTs without detectable traces of
m-SWNTs and impurities, within our detection limits.
Ultracentrifugation (optimum conditions 250,000~g for 60~min) is
the key technique that lead us to the selective extraction of
s-SWNTs from nanotubes powders with PFO as extracting agent in
toluene solution. Note that the centrifugation conditions
(9,000~g, 3~min) described by Nish \textit{et al.} are inadequate
to remove traces of m-SWNTS. \cite{Nish-NN} The present
achievement is extremely importance for both basic and applied
research.

Next, prototypical FET consisting of random network of SWNTs were
fabricated by spin-coating the sample S on Si/SiO$_2$ wafer with
pre-patterned drain and source electrodes. Remarkably, such device
exhibits very promising characteristics rarely observed
simultaneously. The on-off current is around $10^5$, the
on-current is $I_{ON}=15.9\mu$A, the transconductance is
$G_{m}=1.75\mu$S ($G_{m}=\delta I_{D}/ \delta V_{G}$) and the
threshold slop is $S_{p}$=3.15V/decade ($S_{p}=\delta V_{DS}/
\delta Log(I_{D})$). In addition, $I_{D}$ vs $V_{DS}$ exhibits a
well defined linear regime ($\delta I_{D}/ \delta
V_{DS}=-0.2\mu$A/V) and a saturation regime ($I_{sat}=-7.2\mu$A).
Similar results were observed on 10 devices, which were recorded
using the same device structure but not necessarily the same
density of SWNTs. The histogram in figure \ref{fig2}c shows that 5
devices are in the range $R$=5-5.5 and 3 are in the range
$R$=3.5-4 (with $R$ the on-off ratio defined as
$R=Log_{10}(I_{ON}/I_{OFF})$)). Admittedly, the characteristics of
our devices are relatively scattered, but we believe that this
technical challenge can be surmounted and reproducible FET can be
fabricated.

Interestingly by plotting the data related to sample S and M, the
figure \ref{fig2}c can be divided in 3 areas indicated with
circles. On one side, high $R$ is correlated to low $I_{ON}$ and
$G_{m}$, on the other side low $R$ is correlated to high $I_{ON}$
and $G_{m}$, and in between there is an intermediate situation.
Remarkably, $I_{ON}$ and $G_{m}$ decreases linearly with
increasing $R$. In substance, this result suggest that FET devices
can be fabricated with desired transfer characteristics depending
on the targeted application, by simply tuning the centrifugation
conditions (in other words, the ratio semiconducting/metallic
SWNTs) and the spin-coating conditions (density of SWNTs).

To improve the performances of the FET devices, s-SWNTs in NMP
(similar to sample S) were deposited on pre-defined location and
orientation by dielectrophoresis (DEP). \cite{Nico-APL,
Krupke-NanoLett} This technique leads to a relatively dense
network of s-SWNTs oriented perpendicular to the electrodes
eventually bridging the electrodes with a few nanotube-nanotube
junctions (Fig. \ref{fig3}c). Here again, the device present
relatively good characteristics: $R$=4.8, $I_{ON}$=6.5$\mu$A,
$G_{m}$=1.7$\mu$S, and $S_{p}$=2.98V/decade. Similar results were
observed at least for 3 devices (2 devices with $R$=4.8 and 1
device with $R$=3.6 as shown in figure \ref{fig3}c). The estimated
field effect hole mobility is larger than $\mu$=2~cm$^{2}$/vs. The
latter was calculated using established equations \cite{Fukao,
Snow-APL, brutting} with the following parameters $L=5~\mu$m
(source-drain gap), $W=60~\mu$m (channel width), $t_{ox}$=~500nm
(gate thickness) and $\epsilon=34.5~pF/m$ (permittivity of silicon
dioxide). If we normalize $I_{ON}$ and $G_{m}$ to the width of the
electrode $W$, than the performances of the FET made by DEP are
exceeding (by a factor 400~$\mu$m/60~$\mu$m) those made by
spin-coating method.

How our technique and the performances of our FET compare with
previous reports? Fabrication of FET based on individual s-SWNTs
are technically challenging, and methods involving selective
break-down of m-SWNT occasionally damage the entire device.
\cite{Fukao, Snow-APL} In contrast, deposition of already prepared
s-SWNT processable solutions are compatible with FET and TFT
technology. The performances of our device are comparable to those
made using density gradient technique, larger than the
state-of-the-art solution processable polymers and organic
molecules, but lower than amorphous Si and vacuum deposited
Pentacene. \cite{Review-FET, brutting}

In conclusion, we have demonstrated the selective extract s-SWNTs,
without detectable traces of m-SWNTs and impurities, from carbon
nanotube powders using PFO as extracting agent in toluene assisted
in particular by ultracentrifugation (250,000g). We anticipate
that this method will lead to pure s-SWNTs produced in sizable
quantities and formulated as a printable "ink". We have also
demonstrated that s-SWNTs processable solutions can realize
high-performances p-type FET with very promising characteristics:
the on-off current ratio around $10^5$, $I_{ON}$ around 10~$\mu$A
and the estimated hole mobility larger than 2~cm$^{2}$/vs. We
believe that the present achievement pave the way to the future
development of FET and TFT based on s-SWNT. Admittedly, further
study and optimization of both the materials and the devices are
still necessary to met the industrial requirements.

\acknowledgments We thank all the members of the Nano-Carbon Team
(AIST). N. Izard thanks the Japan Society for the Promotion of
Science for financial support.

%Bibliography

\end{document}